\documentclass{emulateapj}\usepackage{apjfonts}

\newcommand{\bNp}{\textit{N}$'$}

\newcommand{\NaI}{\ion{Na}{1}}
\slugcomment{ApJ accepted}

\newcommand{\Model}[1]{Model \textsc{#1}}

\begin{document}

\title{A Ring of Warm Dust in the HD 32297 Debris Disk}
\author{Michael P. Fitzgerald\altaffilmark{1,2},
Paul G. Kalas\altaffilmark{1,2},
James R. Graham\altaffilmark{1,2}}
\altaffiltext{1}{Department of Astronomy, 601 Campbell Hall, University of California, Berkeley, CA 94720} 
\altaffiltext{2}{National Science Foundation Center for Adaptive Optics, University of California, Santa Cruz, CA 95064}
\email{fitz@astron.berkeley.edu}

\begin{abstract}
We report the detection of a ring of warm dust in the edge-on disk surrounding HD\,32297 with the Gemini-N/MICHELLE mid-infrared imager.  Our \bNp-band image shows elongated structure consistent with the orientation of the scattered-light disk.  The $F_\nu(11.2\,\micron)=49.9\pm2.1$\,mJy flux is significantly above the $28.2\pm0.6$\,mJy photosphere.  Subtraction of the stellar point spread function reveals a bilobed structure with peaks 0\farcs5--0\farcs6 from the star.  The disk is detected out to the sensitivity limit at $\sim 1$\arcsec, and the flux in each lobe is symmetric to within 10\%.
An analysis of the stellar component of the SED suggests a spectral type later than A0, in contrast to commonly cited literature values.
We fit three-dimensional, single-size grain models of an optically thin dust ring to our image and the SED using a Markov chain Monte Carlo algorithm in a Bayesian framework.
The best-fit effective grain sizes are submicron, suggesting the same dust population is responsible for the bulk of the scattered light.  The inner boundary of the warm dust is located $0$\farcs5--0\farcs7 ($\sim 65$\,AU) from the star, which is approximately cospatial with the outer boundary of the scattered-light asymmetry inward of 0\farcs5.  If the grains responsible for the asymmetry reside inward of the warm ring, they must have properties that differ from those that produce the bulk of the \bNp-band emission.  The addition of a separate component of larger, cooler grains that provide a portion of the 60\,\micron\ flux improves both the fidelity of the model fit and consistency with the slopes of the scattered-light brightness profiles.  The models indicate an outer boundary of the small grain population beyond 1\farcs8 ($\sim 210$\,AU).
Although the interpretation of the system is uncertain due to the unknown stellar age, previous indirect estimates ($\sim 30$\,Myr) indicate the dust is composed of debris.  The peak vertical optical depths in our models ($\sim 0.3$--$1\times 10^{-2}$) imply that grain-grain collisions likely play a significant role in dust dynamics and evolution.  Submicron grains can survive radiation pressure blow-out if they are icy and porous.  Similarly, the inferred warm temperatures (130--200\,K) suggest that ice sublimation may play a role in truncating the inner disk.
\end{abstract}


\section{INTRODUCTION}\label{sec:intro}

The attrition of the primitive remnants of solar system formation replenishes the dust in circumstellar debris disks.
The scattered light and thermal emission from this dust provide a window into the physical processes governing the evolution of solid material around normal stars, at a time after they have shed their primordial gas and dust envelopes and are transitioning to more mature, nearly dust-free systems~\citep[e.g.][]{backman&paresce93, meyer_etal07}.

HD\,32297~\citep[$d=112^{+15}_{-12}$\,pc;][]{perryman_etal97} is a main-sequence star with a recently discovered circumstellar disk.  \citet{schneider_etal05} first resolved the dust in near-IR scattered light with \textit{HST}/NICMOS (\textit{F110W}).  They detected the near-edge-on inner disk out to 3\farcs3 ($\sim$400\,AU), and found a brightness asymmetry inward of 0\farcs5 and a break in the surface brightness profile at 1\farcs7.  Optical confirmation of the scattered-light disk by~\citet{kalas05} followed, revealing an extended, asymmetric outer disk extending to 15\arcsec\ ($\sim$1680\,AU) and suggesting a blue-scattering \textit{R}-[F110W] dust color.
\citet{redfield07} detected the gaseous component of the disk, finding that HD\,32297 exhibits the strongest \NaI\ absorption of any nearby main-sequence debris disk system.

We report our discovery of spatially resolved emission arising from warm dust surrounding HD\,32297~(\S\ref{sec:observations}), and note the independent discovery of resolved thermal emission by~\citet{moerchen_etal07b}.  We examine the morphology of the observed dust emission by removing the direct stellar contribution and determine that the residual structure suggests a model of thermal emission from an optically thin ring.  We fit such a model to the data and explore the allowed parameter distribution~(\S\ref{sec:analysis}).  To conclude, we consider the implications of these findings in the context of physical processes shaping the disk's physical structure~(\S\ref{sec:discussion}).

\section{OBSERVATIONS}\label{sec:observations}

We observed HD\,32297 and HD\,20893 with MICHELLE on the Gemini-N telescope on the night of 2006 September 19 (GN-2006B-C-12).  HD\,20893 serves as the photometric calibrator and point spread function (PSF) reference.  We imaged these stars in the \bNp\ filter ($\lambda_0=11.2$\,\micron, $\Delta\lambda=2.4$\,\micron) in a chop-nod sequence with a 15\arcsec\ throw.  Observations were chopped and nodded at a position angle (PA) of 120\degr, chosen to be roughly perpendicular to the scattered light disk ($\mathrm{PA}=47\fdg6\pm 1\degr$; G. Schneider, 2005, private communication).  Image quality was good (\bNp-band resolution\,$\sim$\,0\farcs 3), though terrestrial cirrus contributed to a varying background level which sometimes saturated the detector.  The ABBA chop-nod sequences which contained saturated frames were discarded.
Of the 188\,s (23.5\,s) of guided, on-source integration time for HD\,32297 (HD\,20893), 94.0\,s (23.5\,s) was used in the final analysis.
Each ABBA sequence was processed to remove both the sky and instrumental backgrounds via a double difference.  The central, guided images resulting from each ABBA double difference were registered and stacked.

Photometric calibration was performed with HD\,20893, which is a~\citet{cohen_etal99} standard with a zero-airmass mean \bNp-band flux density of $4.29\pm0.14$\,Jy at an isophotal wavelength of 11.23\,\micron.  The mean airmass during the HD\,32297 exposures was $\sim$11\% greater than during the HD\,20893 exposures, and we assumed an extinction of 0.172\,mag\,AM$^{-1}$ (corresponding to a mean $N$-band [$\lambda_0=10.0$\,\micron, $\Delta\lambda=5$\,\micron] Mauna Kea extinction;~\citealp{krisciunas_etal87}) when correcting the measured HD\,32297 fluxes.

\section{RESULTS \& ANALYSIS}\label{sec:analysis}

\begin{figure*}
\plotone{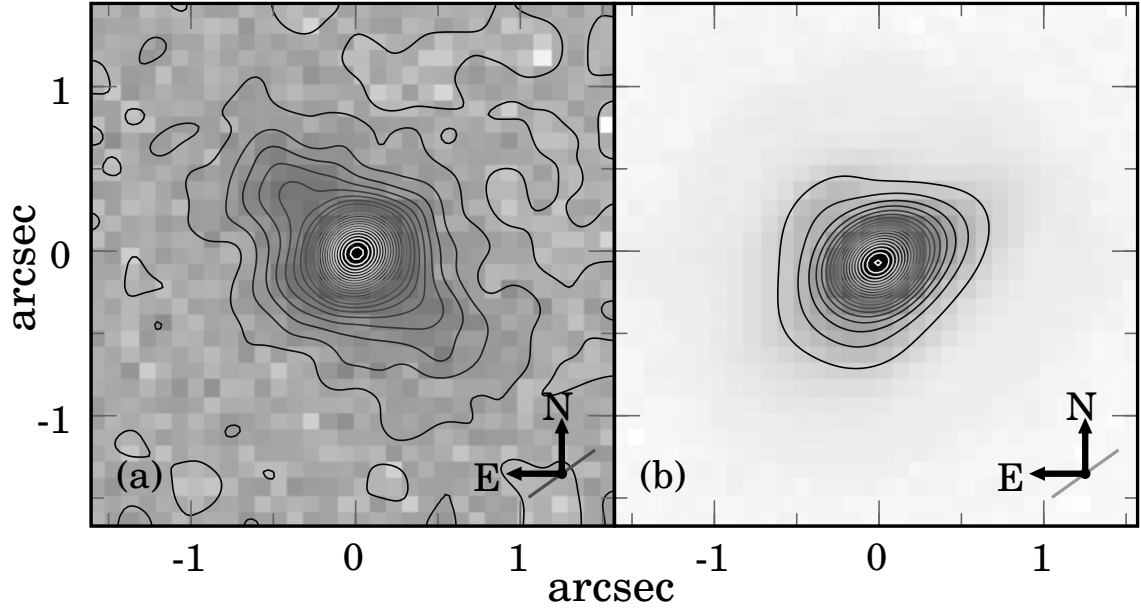}
\caption{
Panel \textit{(a)} displays the result of our Gemini-N/MICHELLE imaging of the HD\,32297 debris disk (\bNp\ band), showing elongated emission around the star. Contours are spaced according to the 1-$\sigma$ background noise level.  The elongation PA is consistent with the scattered light disk.  The chop direction is indicated by the grey line on the compass.
Panel \textit{(b)} shows the image of the reference star HD\,20893.  The contour spacing is selected to match those in \textit{(a)} relative to the stellar flux.
}\label{fig:images}
\end{figure*}

\begin{figure*}
\plotone{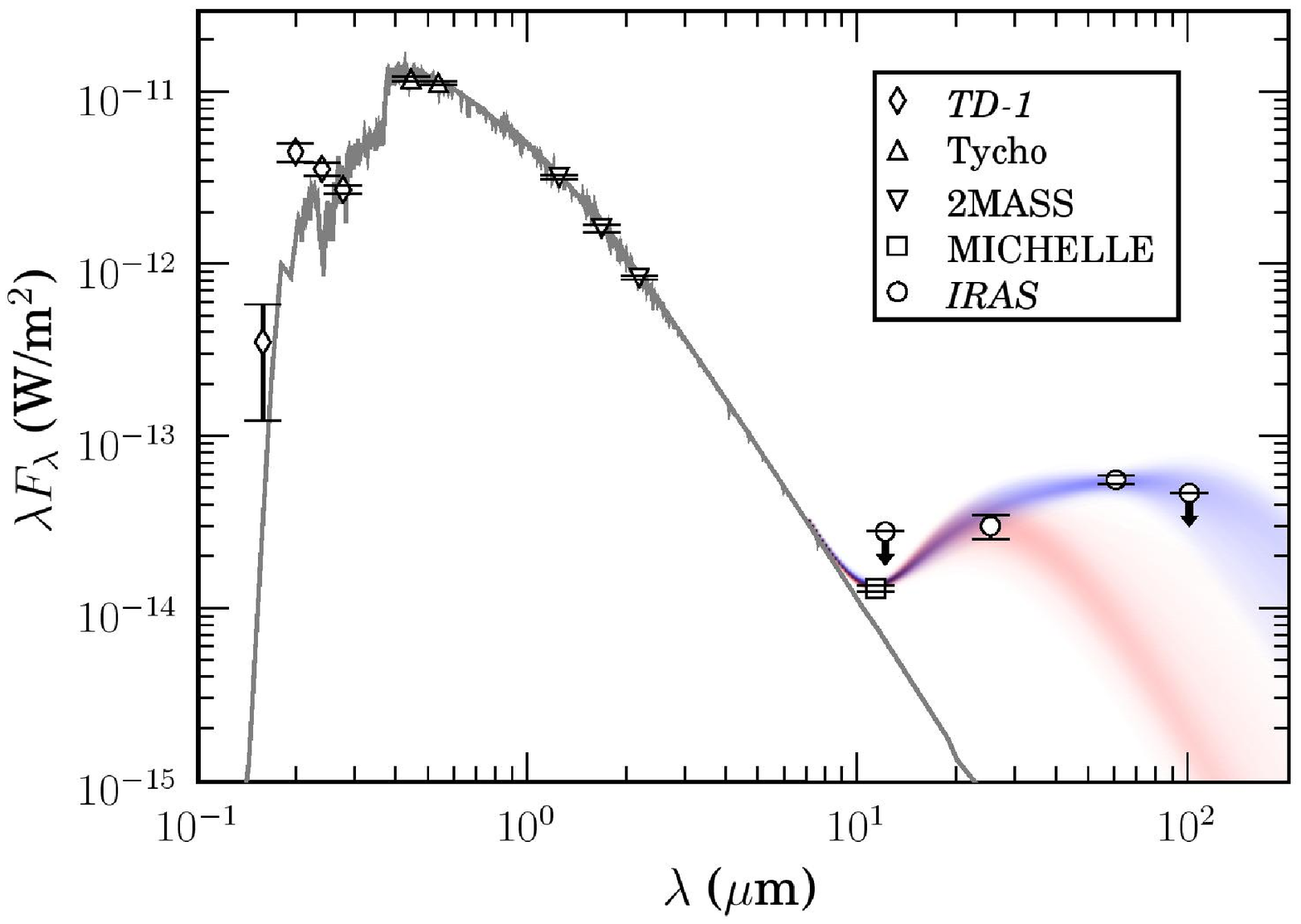}
\caption{The SED of the HD\,32297 system.  \textit{IRAS} photometry is from~\citet{moor_etal06} (see also~\citealt{silverstone00}).  Near-IR data is from 2MASS~\citep{skrutskie_etal06}, optical data from Tycho-2~\citep{hog_etal00}, and UV data from \textit{TD-1}~\citep{thompson_etal78}.  A 7600\,K \textsc{NextGen} model photosphere~\citep{hauschildt_etal99} with zero foreground extinction was fit to the optical-NIR photometry.  The range of allowed dust emission from \Model{I} is shown in blue, while that of \Model{II} is shown in red~(\S\ref{subsec:modeling}).  \Model{I} fits for the image and all SED points, while \Model{II} treats the 60\,\micron\ measurement as an upper limit, effectively allowing for a separate population of cool dust.}\label{fig:SED}
\end{figure*}

The resulting images are shown in Figure~\ref{fig:images}.  The image of HD\,32297 exhibits extension consistent with the direction of the scattered light disk.  The extended emission is detected out to the sensitivity limit at $\sim 1$\arcsec.
With a 1\farcs4 radius aperture and an encircled energy correction derived from the image of HD\,20893, we measured the total flux of the star and disk to be $49.9\pm2.1$\,mJy.  The uncertainty includes the contribution from both the background noise and the zero point uncertainty derived from HD\,20893.  As the filter is relatively narrow, a color-correction to this monochromatic flux density was not considered necessary.  We place the measurement of total flux in relation to the known SED in Figure~\ref{fig:SED}.

\subsection{Stellar Properties}\label{subsec:star}
In order to characterize the dust emission, we must (1) estimate the fraction of the observed \bNp\ flux attributable to the star, and (2) estimate the stellar luminosity, which affects the temperature balance of the grains.
To these ends, we characterize the star by modeling the optical and near-IR portion of the SED,
given by Tycho-2 and 2MASS photometry~\citep{hog_etal00, skrutskie_etal06}.  For model comparison, we synthesized photometry with the stellar atmospheres of the \textsc{NextGen} grid~\citep{hauschildt_etal99}.  The scaling for the resulting photospheric fluxes, parameterized by $\xi\equiv(R_*/d)^2$, was least-squares fit to match the synthetic photometry to the optical/near-IR data.
The choice of model atmosphere parameters is unclear at first glance, since the spectral type of HD\,32297 is inconsistent in the literature.  It was classified as A0 by~\citeauthor{cannon&pickering93} in the Henry Draper Catalog, while it specified as A5 in the AGK3 Catalog~\citep{heckmann75}.
Assuming no interstellar extinction, $\log g=4.5$, and $[\mathrm{Fe}/\mathrm{H}]=0$, we found that a $T_\mathrm{eff}=9600$\,K spectrum (appropriate for A0V) is clearly rejected by the photometry ($\chi^2_\nu=200$), while a 7600\,K \textsc{NextGen} model fit the data best ($\chi^2_\nu=2.5$).  The resulting best-fit model is shown in Figure~\ref{fig:SED}.

Can interstellar reddening bias our inferred $T_\mathrm{eff}$?  At $\sim$112\,pc, the star may be outside the relatively dust-free Local Bubble, which~\citet{redfield07} notes is expected to extend to $\sim$90\,pc in this direction~\citep{lallement_etal03}.  As an empirical test, we fit for the extinction (parameterized by $A_V$) using the $A_\lambda/A_V$ law of~\citet{fitzpatrick04} and assuming $R_V=3.1$.  The best-fit model, fixing $T_\mathrm{eff}=9600$\,K, gives $A_V=0.72\pm0.02$\,mag (formal error) and is rejected with $\chi^2_\nu=8.5$.\footnote{With the addition of $A_V$ as a parameter, the number of degrees of freedom $\nu$ has decreased relative to previous fits.}  When adjusting $T_\mathrm{eff}$, the 7600\,K model again minimizes $\chi^2$, with a best-fit $A_V=0.026\pm0.024$\,mag at this temperature.  Thus, we empirically determine that the observations are consistent with zero interstellar extinction.

As an alternative approach to inferring the amount of interstellar dust toward the star, we consider the measurement of the interstellar component of \NaI\ absorption by~\citet{redfield07}.  He measures a column density of $N_\mathrm{NaI}\sim 10^{12}$\,cm$^{-2}$.  There is considerable scatter in the correlation between $N_\mathrm{NaI}$ and \textit{E}(\textit{B}-\textit{V}), and assuming $R_V=3.1$, values of $A_V=0.03$ to 0.3\, mag are reasonable~\citep[Fig.~4d of][]{hobbs74}.  Since this is consistent with a small amount of extinction, we do not alter our conclusions.

We estimate the flux scaling $\xi=(7.22\pm0.14)\times 10^{-20}$ from the best-fit zero-extinction model (after scaling the formal uncertainty by $\sqrt{\chi^2_\nu}$).  Using this quantity, we calculate the monochromatic flux density $F_\nu(11.23\,\micron)=28.2\pm 0.6$\,mJy.
The uncertainties in our estimates of the stellar radius and luminosity are dominated by distance errors, as
\begin{equation}
L_*=4\pi \xi d^2\sigma T_\mathrm{eff}^4.\label{eq:stellar_lum}
\end{equation}
Taking the best estimates $\xi=7.22\times 10^{-20}$ and $d=112$\,pc, we find $R_*=1.34$\,$R_\sun$ and $L_*=5.4$\,$L_\sun$. 
Together, these suggest the literature value of A5 is closer to the ``true'' classification, though the cool temperature we estimate suggests even later A subclasses.  As noted by~\citet{schneider_etal05}, the star lies near the bottom of the A star main sequence in a color-magnitude diagram, a property in common with other young debris disk systems~\citep{jura_etal98}.  Figure 2 of~\citet{moor_etal06} shows that HD\,32297 is relatively under-luminous and is positioned toward the red end of the local A star sequence, consistent with our results.  A modern high-resolution spectrum can settle the debate over the correct stellar classification.

Finally, we note the ultraviolet measurements from the \textit{TD-1} satellite~\citep{thompson_etal78} shown in Figure~\ref{fig:SED}.  We did not use these data in our photometric fit.  The \textit{F1965} and \textit{F2356} bands ($\lambda=196.5$ and 235.6\,nm, respectively) show excess above the model photosphere.  These may be due to line emission --- such as \ion{Si}{3]} and \ion{C}{3]} in the former band, and \ion{Fe}{2} in the latter --- similar to what is seen in T\,Tauri stars~\citep[e.g.][]{lamzin00b, valenti_etal03}.  In such stars, the intercombinational lines are likely associated with accretion rather than the chromosphere~\citep{lamzin00a}.  The UV excesses in HD\,32297, along with the recent evidence for a gas disk~\citep{redfield07}, lend indirect support to the contention that this is a young stellar system.

\subsection{PSF Subtraction}\label{subsec:psfsub}
\begin{figure}
\plotone{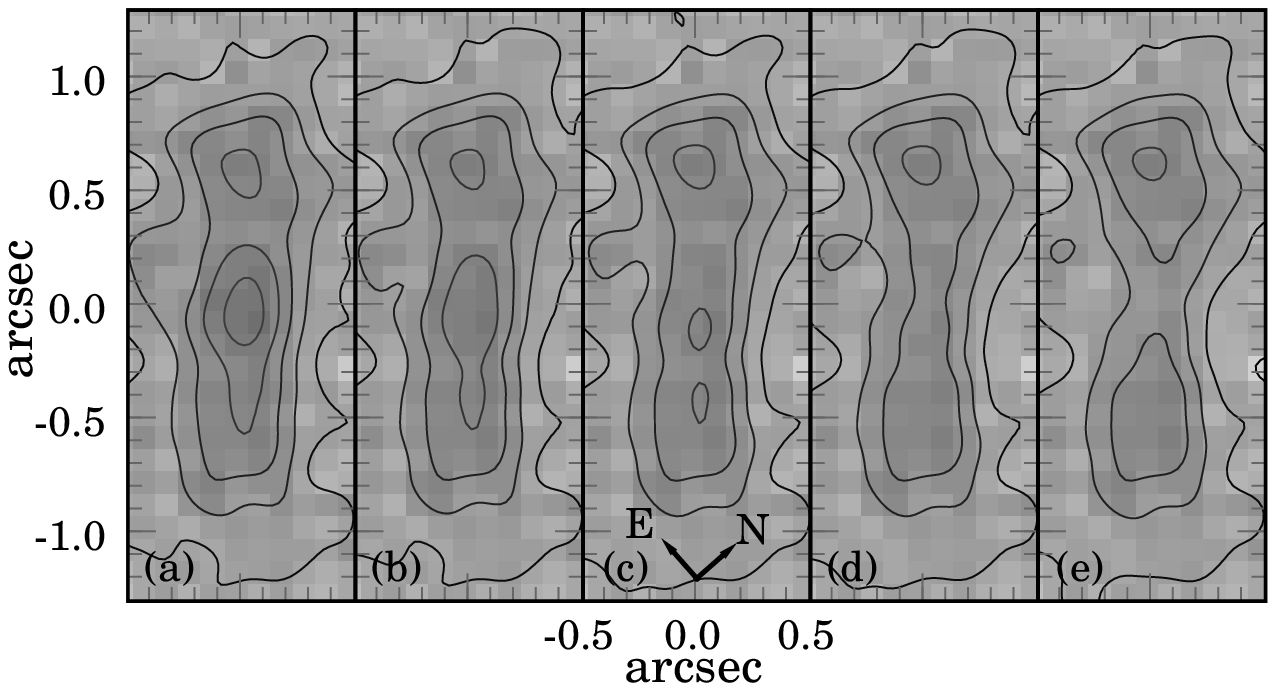}
\caption{Images of the disk with the stellar PSF subtracted for various values of the stellar flux estimate.  With $F_\nu^0\equiv 28.2$\,mJy and $\sigma\equiv 1.1$\,mJy (which includes both error in our flux estimate,~\S\ref{subsec:star}, and \bNp\ calibration uncertainty,~\S\ref{sec:observations}), panels \textit{(a)}--\textit{(e)} depict subtractions of a central source with $F_\nu=\left\{F_\nu^0-2\sigma, F_\nu^0-\sigma, F_\nu^0, F_\nu^0+\sigma, F_\nu^0+2\sigma\right\}$, respectively.  As in Fig.~\ref{fig:images}, contours are spaced according to the 1-$\sigma$ background noise level.  The bilobed residuals suggest that the system may be modeled as a central star with an optically thin ring of warm material.}\label{fig:psfsub}
\end{figure}

Using the range of $\xi$ estimated in the previous section, we scaled the image of HD\,20893 and subtracted it from the that of HD\,32297.
After PSF subtraction, we find $21.6\pm 2.2$\,mJy in non-color-corrected residual flux
(Fig.~\ref{fig:psfsub}).  The residuals suggest the observed image (Fig.~\ref{fig:images}a) is a composite of a bilobed structure and an unresolved central source, whose flux can be fully accounted for by the star.  The lobes peak at offsets of 0\farcs5--0\farcs6 from the star, corresponding to a radius of $\sim$60\,AU.  This suggests an inner edge to the population of grains with optically thin \bNp-band emission --- both hotter dust closer to the star and the case of optically thick material would fill in the emission at smaller projected separations.

We quantified the degree of asymmetry between the disk lobes with aperture photometry of the PSF-subtracted disk image (Fig.~\ref{fig:psfsub}c).  We used 0\farcs6 square apertures placed on each lobe, with centers 0\farcs6 from the stellar centroid.  Neglecting an aperture correction and calibration uncertainty, we measured a NE-SW flux difference of $0.17\pm0.36$\,mJy.  As a fraction of the total flux in these apertures, the difference is $(2.0\pm4.3)\times 10^{-2}$.  We conclude that the disk emission is consistent with symmetry.

\subsection{Modeling}\label{subsec:modeling}

\subsubsection{Model Construction}\label{subsubsec:model_constr}
We adopt the model for thermal emission from optically thin dust rings developed by~\citet{backman_etal92}, which assumes particles of effective size $\lambda_0$ that radiate as modified blackbodies with emission efficiencies $\epsilon_\nu$.
We model the three-dimensional emission of the dust $j_\nu(\mathbf{x})$ by evaluating
\begin{eqnarray}
\alpha_\nu(\mathbf{x}) &=&\left\{
\begin{array}{ll}
\tau_0 \left(\frac{\varpi}{\varpi_0}\right)^\gamma f\left[z,h_0\left(\frac{\varpi}{\varpi_0}\right)^\eta\right] & \mbox{for $\varpi_0\leq\varpi\leq\varpi_1$,} \\
0 & \mbox{otherwise},
\end{array}\right. \label{eq:alpha_nu}\\
\epsilon_\nu &=&\left\{
\begin{array}{ll}
\lambda_0/\lambda & \mbox{if $\lambda>\lambda_0$,} \\
1 & \mbox{otherwise},
\end{array}\right. \label{eq:emissivity} \\
T(r) &=& 468 \left(\frac{L_*/L_\sun}{\lambda_0/1\,\micron}\right)^{1/5} \left(\frac{r}{1\,\mathrm{AU}}\right)^{-2/5} \mathrm{K},\label{eq:temp} \\
j_\nu(\mathbf{x}) &=& \alpha_\nu(\mathbf{x}) \epsilon_\nu B_\nu\left[T(r)\right].\label{eq:j_nu}
\end{eqnarray}
Here, $\alpha_\nu$ is the absorption coefficient, and the function $f(z,h)$ describes the vertical distribution of dust and has scale height $h$.  The fiducial values for the vertical optical depth to absorption $\tau_0$ and scale height $h_0$ are set at the inner edge of the dust annulus, $\varpi_0$.  We use cylindrical coordinates for positions $\mathbf{x}$, with $r^2=\varpi^2+z^2$, and assume azimuthal symmetry.

Some assumptions of grain properties are implicit in Equations~\ref{eq:alpha_nu}--\ref{eq:j_nu}.  The disk is assumed to be optically thin at all radii.  The particles are efficient absorbers, but inefficient emitters (Eq.~\ref{eq:emissivity}).  \citet{backman_etal92} discuss the relationship between $\lambda_0$ and the particle absorption efficiency and size distribution in their Appendix D.  They argue that for ``dirty ice'' dielectric constants and $dn(a)\propto a^{-7/2} da$ collisional-cascade size distribution~\citep{dohnanyi69}, the minimum grain size is an uncertain $a_\mathrm{min}\sim\lambda_0/6$.  With these assumptions, the model contains only minimal information regarding grain characteristics.  While we ignore the effects of grain composition with this approach, the absence of evidence for grain mineralogy (e.g. mid-IR spectroscopy) renders consideration of various compositions premature.  Future work using more advanced models with explicit dependence on grain sizes and mineralogy may produce significant differences from our results.

For simplicity, we assume the vertical density distribution $f$ is Gaussian, with width $\sigma=h_0=10\left(d/112\,\mathrm{pc}\right)$\,AU fixed at all radii ($\eta=0$).  Because we do not measure a significant brightness asymmetry~(\S\ref{subsec:psfsub}), we assume the disk is spatially centered on the star.
The distance $d$ is a `nuisance' parameter not directly related to the disk architecture, and in the interest of independence we define $\theta_0\equiv\varpi_0/d$ (likewise for $\theta_1$).
The remaining free parameters are $\theta_0$, $\theta_1$, $\lambda_0$, vertical optical depth to absorption $\tau_0\equiv\tau_\perp^\mathrm{abs}(\varpi_0)$, (subpixel) stellar centroid $\mathbf{x}_*$, stellar flux (parameterized by $\xi$;~\S\ref{subsec:star}), disk inclination $i$, and PA.  The stellar luminosity is given by Eq.~\ref{eq:stellar_lum}.  There are 11 degrees of freedom within the set of model parameters $\Theta = \left\{d, \mathbf{x}_*, \mathrm{PA}, i, \xi, \gamma, \theta_0, \theta_1, \tau_0, \lambda_0, h_0, \eta\right\}$.

We construct the density distribution on a three-dimensional grid, with spatial resolution of $0\farcs05=5.6\left(d/112\,\mathrm{pc}\right)\,\mathrm{AU}$, which is half the spatial sampling of the MICHELLE detector (0\farcs1\,pix$^{-1}$).  The emission coefficient $j_\nu$ (Eq.~\ref{eq:emissivity}) is sampled at the grid points and numerically integrated along each line of sight.
As in the simple subtraction in~\S\ref{subsec:psfsub}, the PSF is derived from the image of HD\,20893 (Fig.~\ref{fig:images}b).  We upsample the PSF to match the grid resolution, shift it according to $\mathbf{x}_*$, and convolve it with the disk model's integrated emission.  These procedures take place in the Fourier domain with their Fourier-equivalent operations via the convolution and shift theorems.
We transform back to the image domain and bin the result to the resolution of the instrument (a factor of 2 in each dimension).

\subsubsection{Fitting Process}\label{subsubsec:model_fit}
Given the model disk, how do we characterize the range of model parameters allowed by the data?
As we have only limited information (e.g. the thermal emission resolved in a single band), we expect that degeneracies will exist between some parameters (such as $\lambda_0$ and $\tau_0$).
Further, because Eqs.~\ref{eq:stellar_lum}--\ref{eq:j_nu} contain significant nonlinearities, we expect non-Gaussianity in the joint distribution of allowed parameters.
Finally, this problem contains nuisance parameters (e.g. $d$) which are unimportant when considering the physical architecture of the disk.
These characteristics, combined with a desire to incorporate prior knowledge, suggest a Bayesian approach to model fitting.  The range of disk architectures allowed by our simple model can be described by the joint posterior distribution of model parameters, namely
\begin{equation}
p(\Theta|\mathcal{D},I) \propto p(\mathcal{D},\Theta|I)=p(\mathcal{D}|\Theta,I)p(\Theta|I).\label{eq:posterior}
\end{equation}
The posterior distribution of parameters is proportional to the likelihood, $p(\mathcal{D}|\Theta,I)$, times the prior, $p(\Theta|I)$.  Here, $\mathcal{D}$ represents the observational data and $I$ represents our background information.  In this framework, the best-fit model is one whose parameters maximize the posterior distribution (so-called \textit{maximum a posteriori} fitting).
The Bayesian approach has previously been used in modeling mid-IR images of debris disks~\citep[e.g.][]{koerner_etal98, wahhaj_etal03}.  A more general comparison of Bayesian and frequentist methods in the context of common astronomical problems is given in~\citet{loredo92}.

Brute-force characterization of this distribution is not computationally efficient.  The large number of model parameters ensures that the posterior has a relatively high dimensionality, and grid-based schemes for mapping the posterior require many evaluations of the model.  Instead, we turn to Markov chain Monte Carlo (MCMC) methods to more efficiently explore the parameters' joint posterior distribution.  We note that these advanced statistical techniques stand in contrast to the relative simplicity of our model; our goal of measuring the ranges of allowed disk architectures motivates the expenditure of computational resources in the exploration of parameter space rather than the complexity of the underlying disk model.

The MCMC framework uses a Markov chain, consisting of a series of states $\{\Theta_n\}$, to explore the parameter space and sample the posterior.  A requirement of MCMC methods is an algorithm for randomly choosing states in the Markov chain.  Given a state $\Theta_n$, the Metropolis-Hastings algorithm provides a general-purpose method for randomly choosing the subsequent state $\Theta_{n+1}$ based on a candidate transition function (commonly Gaussian).  Often this algorithm is used in conjunction with the Gibbs sampler, which provides a prescription for choosing which of the model variables to change in each transition.  With the Metropolis-Hasting algorithm choosing states, the chain will eventually reach convergence.  A key property of converged chains is that the distribution of states approximates the posterior distribution.
The use of MCMC in astrophysical data analysis has grown in recent years; a summary of this technique and an application to quantify uncertainty in orbit fitting is given by~\citet{ford05}.

When addressing questions about the physical properties of the disk, we are sometimes interested only in the range of a subset of parameters allowed by the model.  In the Bayesian framework, the distribution of a parameter (or joint distribution of a subset of parameters) is obtained through marginalization.  The posterior is integrated over the variables not of interest.  For example, to obtain the joint marginal distribution of $\tau_0$ and $\lambda_0$, we integrate over the other variables $\Theta'=\left\{d,\mathbf{x}_*,\mathrm{PA},i,\xi,\gamma,\theta_0,\theta_1,h_0,\eta\right\}$,
\begin{equation}
p(\tau_0, \lambda_0|\mathcal{D},I) = \int p(\Theta|\mathcal{D},I) d\Theta'.
\end{equation}
Having obtained an MCMC chain, this integral is trivial --- the joint marginal distribution is simply the joint distribution of the variables' samples, in this case the distribution of $\{\tau_{0,n},\lambda_{0,n}\}$.

Having chosen the sampling scheme, we must define the likelihood and prior distributions for our model parameters.  We calculate the likelihood by subtracting the model image from the observed emission presented in Fig.~\ref{fig:images}a.  We compute $\chi^2_\mathrm{im}$ for the image fit using the background noise level, and we add the $\chi^2_\mathrm{SED}$ arising from the \textit{IRAS} SED measurements (including upper limits).  The likelihood is then $p(\mathcal{D}|\Theta,I)\propto\exp\left[-(\chi^2_\mathrm{im}+\chi^2_\mathrm{SED})/2\right]$.

We have several priors whose product constitutes $p(\Theta|I)$.  We use normal priors for the \textit{Hipparcos} parallax ($8.92\pm1.05$\,mas; a proxy for $d$) and the stellar flux factor $\xi$~(\S\ref{subsec:star}).  We also apply a log-uniform prior to $\lambda_0$, with limits of 1\,nm--1\,mm, as well as a uniform prior on $\gamma$, ranging from -4 to 4.  The prior for the outer disk extent $\theta_1$ is taken to be uniform from $\theta_0$ to 200\arcsec.
With these likelihood and prior functions, we ran a total of six chains using the Metropolis-Hastings algorithm with the Gibbs sampler.  Each chain contained $10^4$ samples.  The first $10^3$ samples were discarded, as the candidate transition function proposal variance for each parameter was adjusted during this ``burn-in'' period.
We expect the Markov chains have converged because no gross deviations exist when comparing the marginal distributions of each parameter across all chains.

In general, the convergence rate and outcome of fitting processes are dependent on starting conditions.  Rather than input hand-picked parameters to the MCMC algorithm, we first apply a genetic algorithm (for global optimization) followed by a Levenberg-Marquardt least-squares fit (for local optimization) to find a model of the image with suitable parameters.  During this process, the (nuisance) distance is fixed at 112\,pc.
Examination of the marginal distributions of each parameter shows that the posterior distribution in the region of the global maximum is smooth.  As a check for other maxima, we drew 500 randomly distributed samples of $\Theta$ over a wide distribution, and performed a least-squares fit with each sample as the starting condition.  All fits converged on the same maximum, which suggests the absence of lesser maxima in the posterior distribution.

\subsubsection{Results}\label{subsubsec:model_results}
Using the ring model presented in~\S\ref{subsubsec:model_constr}, we fit the \bNp-band image, and in the SED the 25 and 60\,\micron\ points while respecting the 12 and 100\,\micron\ upper limits (\Model{I}).  Confidence intervals for the marginal posterior parameter distributions are given in Table~\ref{tab:parms}.
We are able to obtain a reasonable fit to the image, though the model tends to overpredict the 25 and 100\,\micron\ fluxes (Fig.~\ref{fig:SED}).
With the parameters that maximize the posterior, the $\chi^2$ for the image is 912 (with 32$^2$ pixels), while the $\chi^2$ for the \textit{IRAS} SED is 3.3 (with 2 data points and 2 upper limits).

\begin{deluxetable}{cccl}
\tablewidth{0pc}
\tablecaption{Best-Fit Model Parameters\label{tab:parms}}
\tablecolumns{4}
\tablehead{
\colhead{parameter} & \colhead{\Model{I}} & \colhead{\Model{II}} & \colhead{Notes}
}
\startdata
PA                                           & $44\fdg5^{+2.6}_{-2.5}$     & $44\fdg5\pm2\fdg5$          & \nodata \\
$i$                                          & $90\fdg0\pm4\fdg7$          & $90\fdg0\pm4\fdg8$          & \nodata \\
$\xi$                                        & $1.02\pm0.03$               & $1.03\pm0.03$               & $\times7.22\times10^{-20}$ \\
$\gamma$                                     & $0.07\pm0.22$               & $-2.1^{+2.2}_{-1.7}$        & \tablenotemark{a} \\
$\theta_0$                                   & $0\farcs54^{+0.05}_{-0.07}$ & $0\farcs65^{+0.06}_{-0.09}$ & \nodata \\
$\theta_1$                                   & $>33\arcsec$                & $>1\farcs8$                 & \nodata \\
$\log_{10}\left(\tau_0\right)$               & $-2.57^{+0.24}_{-0.30}$     & $-1.94^{+0.27}_{-0.49}$     & \nodata \\
$\log_{10}\left(\lambda_0/1\,\micron\right)$ & $-0.65^{+0.21}_{-0.27}$     & $-0.41^{+0.15}_{-0.22}$     & \nodata \\
\sidehead{Dependent or fixed parameters:}
$h_0$                                        & \multicolumn{2}{c}{10($d/112$\,pc)\,AU}                   & \nodata \\
$\eta$                                       & \multicolumn{2}{c}{0}                                     & \nodata \\
$\varpi_0$                                   & $61^{+20}_{-13}$\,AU        & $73^{+23}_{-16}$\,AU        & \nodata \\
$\varpi_1$                                   & $>3700$\,AU                 & $>210$\,AU                  & \nodata \\
$T(\varpi_0)$                                & $171^{+31}_{-19}$\,K        & $142^{+21}_{-11}$\,K        & \nodata \\
\enddata
\tablenotetext{a}{\,Lower-limit in \Model{II} set by prior.}
\tablecomments{95\% confidence intervals for marginal posterior distributions.  PA, position angle; $i$, inclination; $\xi=(R_*/d)^2$, stellar flux parameter; $\gamma$, surface density power law index; $\theta_{0,1}$, disk inner/outer edges, angular units; $\tau_0$ vertical optical depth to absorption at inner edge; $\lambda_0$, effective grain size; $h_0$, scale height at inner edge; $\eta$, scale height power law (cf. Eq.~\ref{eq:alpha_nu}); $\varpi_{0,1}$, inner/outer edges, spatial units; $T(\varpi_0)$, dust temperature at inner edge.}
\end{deluxetable}

The marginal distributions for the position angle and inclination are approximately normal, with $\mathrm{PA}=44\fdg4\pm1\fdg3$ and $i=90\fdg0\pm2\fdg4$.  The PA is marginally inconsistent with the direction of the inner scattered light disk, though this should not be taken as evidence of separate disks since we did not calibrate the detector orientation.  We note that the range of allowed inclinations may be affected by our choice of disk scale height.

The SED data at $\geq 25$\,\micron\ constrain the values of the outer radius $\theta_1$ and density power-law index $\gamma$.  From the marginal parameter distribution, we find $\theta_1\gtrsim30$\arcsec, which is further than the scattered light disk has been detected~\citep[15\arcsec;][]{kalas05}.
The model also requires $\gamma\simeq0$, which places a large amount of cool material in the outer disk.
With $\lambda_0\simeq0.2$\,\micron, giving $a_\mathrm{min}\ll1$\,\micron, we can infer this to be the same population giving a blue \textit{R}-[F110W] color, expected for small grains approaching the Rayleigh regime~\citep{kalas05}.
We note that Rayleigh grains scatter light quasi-isotropically, in the sense that the first moment of the scattering phase function ($g\equiv\langle\cos\theta\rangle$) is zero.  For an edge-on, wedge-shaped disk ($\eta=1$) of a power-law distribution of isotropic scatterers, the midplane surface brightness profile at projected distance $b$ is $\propto b^{\gamma-1}$.  Under these assumptions, \Model{I} predicts a brightness profile power law index of $\sim-1$, which is shallower than the measured indices~\citep[-2.7 to -3.7;][]{schneider_etal05,kalas05}.  This suggests that the long-wavelength SED is not produced by a relatively flat distribution out to large radii, but rather by a separate population of grains.  Similar populations have been invoked for other systems~\citep[e.g. in $\beta$\,Pic and AU\,Mic;][]{backman_etal92, fitzgerald_etal07}.

The large spatial extent of \Model{I} and its inconsistency with the scattered light profile power law suggest that we amend the model with an additional grain population.
One possibility would be to maintain a distribution of small grains, similar to \Model{I}, to produce the \bNp-band image and 25\,\micron\ flux, and add larger, cooler grains that reproduce the 60\,\micron\ flux but contribute little to the shorter-wavelength emission.  This would require additional parameters; modeling a single ring of large grains requires the ring distance and the grains' size and number.
In the interest of computational simplicity, we choose not to explicitly parameterize the larger grains in \Model{II}.  Instead, we implicitly allow for this population by relaxing the SED fit, requiring only that the 60\,\micron\, flux from the grain population producing the spatially resolved emission does not exceed the \textit{IRAS} measurement.  The unmodeled population is presumed to supplement the small grain emission to match the observed 60\,\micron\ flux.  The larger grains are assumed to emit little at 25\,\micron.  In effect, we treat the 60\,\micron\ detection as an upper limit.

After running the MCMC chains with the same procedure as before, we find that \Model{II} reproduces the \bNp-band image and the 25\,\micron\ flux.
With the parameters that maximize the posterior, the $\chi^2$ for the image is 902, while the $\chi^2$ for the \textit{IRAS} SED is 0.4 (with 1 data point and 3 upper limits).
One drawback to our decision to avoid explicitly parameterizing the large grains in this model is that it precludes rigorous comparison of the goodness-of-fit to \Model{I}.\footnote{This is usually done by comparing the ``evidence'' for each model, through the ratio $\int p(\mathcal{D}, \Theta_\mathrm{I}|\mathrm{Model\ I}, I)d\Theta_\mathrm{I} / \int p(\mathcal{D}, \Theta_\mathrm{II}|\mathrm{Model\ II}, I)d\Theta_\mathrm{II}$, with a possible additional factor representing prior preference for one model over the other~\citep[e.g.][Ch.~4]{sivia06}.}  However, we expect that when combined with scattered-light data, this model will be favored because of its ability to include surface density gradients with $\gamma < 0$.

\begin{figure}
\plotone{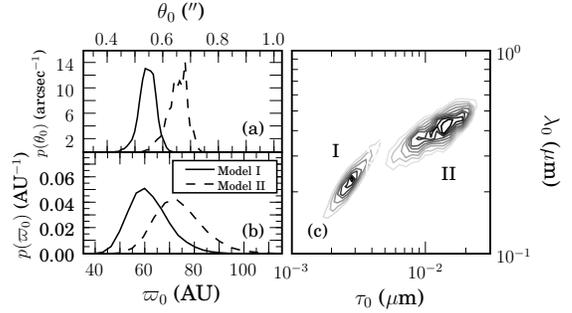}
\caption{Marginal posterior distributions of a subset of model parameters.  Panels \textit{(a)} and \textit{(b)} give the probability density of location of the inner disk edge in angular ($\theta_0$) and physical ($\varpi_0$) units, respectively.
The ordinates of these plots are matched assuming a distance of 112\,pc, and the differences evident between the distributions highlights the value of the joint estimation of all parameters.  Panel \textit{(c)} shows the joint marginal posterior distribution for effective grain size $\lambda_0$ and vertical optical depth to absorption at the inner edge $\tau_0$.}\label{fig:parmdist}
\end{figure}

\Model{II} produces less stringent requirements on $\gamma$ than from \Model{I}, favoring steeper dropoff, though flat distributions are not excluded.
The distributions of the inner radii, in terms of both angular ($\theta_0$) and physical ($\varpi_0$) variables, are given in Figures~\ref{fig:parmdist}a and~\ref{fig:parmdist}b.

To give a basic example of possible parameters for the unmodeled large grain component, we added the flux from a single ring of large grains at 80\,AU to the best-fit \Model{II}.  We found that $10^{29}$\,cm$^2$ of $\lambda_0=85$\,\micron\ grains were able to reproduce to 60\,\micron\ flux without strongly affecting the 25\,\micron\ component from the small grains.  These large grains have a temperature of 50\,K according to Eq.~\ref{eq:temp}.
These large-grain parameters are only a single sample in a range of possibilities, and we note that they are expected to be strongly covariant without additional observations of the far-IR and sub-mm flux.

In both models, the effective grain size $\lambda_0$ and the fiducial optical depth $\tau_0$ are covariant.  This is evident in their marginal joint distributions, shown in Figure~\ref{fig:parmdist}c.  The relaxation of the far-IR SED requirement allows for a broader range for these parameters in \Model{II}.  We stress that joint estimation of parameters is frequently necessary in disk models.  For example, in scattered-light modeling of AU\,Mic, \citet{graham_etal07} revealed a strong degeneracy between scattering asymmetry and the surface density power law, which was broken by the independent constraint from polarization measurements.

In principle, the choice of the prior $p(\Theta|I)$ can affect the distributions of parameters given in Table~\ref{tab:parms}.  A parameter's distribution is sensitive to the prior if it is not dominated by the likelihood term in Eq.~\ref{eq:posterior}.  We explored the sensitivity to choice of prior by comparing the marginal distributions of each parameter with its prior.  In general, both Models \textsc{I} \& \textsc{II} show similar sensitivity to priors.
In each model, the marginal posterior distribution of $\xi$ is roughly Gaussian in shape, however its mean is offset from the mean of the Gaussian prior.  In this case, both the likelihood and prior play a significant role in shaping the posterior distribution.
The marginal posterior distribution of the distance $d$ is completely dominated by the prior derived from the \textit{Hipparcos} parallax.  This is unsurprising, since we did not use a prior for the stellar radius or luminosity that might serve to constrain the distance.
The marginal posterior for the outer radius $\theta_1$ largely follows the flat prior (out to 200\arcsec), though it has an inner cutoff set by the likelihood function.  Thus the lower limits listed in Table~\ref{tab:parms} are relatively insensitive to the prior.
The marginal posterior of $\gamma$ is well-constrained by the likelihood function in \Model{I}, however in \Model{II} the prior affects the posterior.  In the latter model, the distribution shows an upper-limit cutoff near $\gamma=0$ and a peak near $\gamma=-2$.  However, the distribution is smooth down to the lower cutoff set by the prior at $\gamma=-4$.  Therefore the uniform prior for $\gamma$ plays a role in shaping the posterior distribution in \Model{II}, and the likelihood function is only able to exclude values in the upper end of the range.
For all other variables, the likelihood function is sharply peaked compared to the relatively flat priors; therefore their results are insensitive our choice of prior.

\section{DISCUSSION}\label{sec:discussion}

The spatial locations and inferred sizes of the thermally emitting grains are largely consistent with the dust seen in scattered light.
\citet{schneider_etal05} find that the scattered-light disk is symmetric for $0\farcs5<\theta<1\farcs7$, which overlaps with the symmetric mid-IR emission we measure in~\S\ref{subsec:psfsub}.
The inner edge of the warm dust ring ($\theta_0\simeq0$\farcs5--0\farcs7) corresponds to the outer boundary of the scattered-light brightness asymmetry seen inward of 0\farcs5.  A comparison of the NE and SW \textit{F110W} surface brightness profiles shows the SW ansa is $\sim 2$ times brighter than the NE at 0\farcs5.  Integrating the disk light in each ansa from $\theta>0\farcs3$,~\citeauthor{schneider_etal05} find the total NE (SW) emission to be $1.67\pm0.57$\,mJy ($3.14\pm0.57$\,mJy), which corresponds to a NE-SW fractional flux difference of $-0.31\pm0.18$.
In comparison to the scattered light, in the \bNp-band image we do not find an increasing brightness trend for the SW ansa inward of 0\farcs6, nor do we find evidence for asymmetry in this region.  How can the emission appear asymmetric in scattered light, but symmetric in thermal emission?  One possibility is that the grains responsible for the scattered-light asymmetry reside inward of the warm ring.  In this case, they must have properties that differ from the majority of grains in the inner disk, which scatter light in the symmetric component.  If the asymmetrically distributed grains have a significantly higher albedo or larger average size, they may not be apparent in the thermal emission.  Future observations of scattered light at different wavelengths may constrain such changes in grain properties with position.  Another possibility is that the asymmetry is produced by a density enhancement at sufficient distance from the star such that its grains do not produce significant \bNp-band emission relative to the warm ring.

For $\lambda_0\simeq0.1$--0.5\,\micron\ and reasonable assumptions on grain composition and size distribution~(\S\ref{subsubsec:model_constr}), we find $a_\mathrm{min}\sim\lambda_0/6\simeq0.02$--0.1\,\micron.  As noted in~\S\ref{subsubsec:model_results}, in the \textit{R} and \textit{F110W} bands, submicron grains will scatter blue as their sizes approach the Rayleigh regime, consistent with the blue color of dust scattered light inferred by~\citet{kalas05}.  This lends support to the contention that the scattered-light images and thermal image presented here are probing similar populations of grains.  Simultaneous modeling of the scattered light and thermal emission, deferred for future work, has the potential to strongly constrain grain sizes, locations, and composition for the bulk of the inner disk.

These observations are a stepping stone to understanding the nature of the disk and source of its structure.  However, we lack direct evidence for the age of the system, which complicates such analysis.  Based on the stellar distance and galactic space motion, \citet{kalas05} argues for an age $\sim$30\,Myr based on the system's possible association with the Gould Belt or recent star formation in the Taurus-Aurigae association.  As we will show in the following calculations, this age indicates that the grains in the inner disk are likely not primordial and must be replenished.  In the next subsection, we seek to estimate orders of magnitude for processes governing grain production and removal for the sizes and spatial locations in the innermost disk.

\subsection{Grain Dynamics}\label{subsec:dynamics}

Assuming the radiative coupling efficiency averaged over the stellar spectrum $Q_\mathrm{rad}=2$, a grain density of 2\,g\,cm$^{-3}$, and $M_*=1.8$\,$M_\sun$, the fiducial radiation pressure blow-out size is $a_\mathrm{blow}\sim3$\,\micron~\citep{burns_etal79}.  In the absence of forces other than radiation and gravity, grains smaller than this size are removed from the system on the free-fall timescale, $\sim10^4$\,yr.  At first glance, this would imply that production of submicron grains must be extremely rapid.  However, our conversion from $\lambda_0$ to the geometric size $a_\mathrm{min}$ is quite uncertain, and may be too small by a factor of $\sim30$~\citep[if the grains are weakly absorbing and have a size distribution steeper than $dn\propto a^{-3.5}da$;][]{backman_etal92}.
Another possibility, which maintains consistency with submicron scattered-light grains, is that drag forces can increase the residence time of the grains undergoing ejection.
Finally, we note that $a_\mathrm{blow}$ is not a strict lower limit to the steady-state size distribution.  As noted by~\citet{burns_etal79}, very small grains (sizes much less than the peak wavelength of stellar radiation) couple inefficiently to the radiation field, such that $Q_\mathrm{rad}\ll 1$ and the radiation force cannot overcome gravity.  For such small grains, composition and porosity can can play a crucial role in the residence time due to their effects on the optical constants.  For example,~\citet{grigorieva_etal07} show in their Fig.~1 that around an A5V star, silicate grains are blown out regardless of size or porosity.  In contrast, the smallest icy grains can remain bound.  Compact icy grains have a blow-out size of 3\,\micron, whereas $a_\mathrm{blow}$ for grains with 80\% porosity drops to 0.7\,\micron.  Because of their small $Q_\mathrm{rad}$, the radiation pressure on 80\% porous, icy grains smaller than 0.05\,\micron\ is insufficient to overcome gravity.  This cutoff size decreases to $\sim 0.02$\,\micron\ for more compact icy grains.
We conclude that our model grain sizes may be compatible with the steady-state size distribution arising from radiation pressure. and note that this mechanism provides a natural separation between a population of small ($a\lesssim 0.05$\,\micron) grains and a population of grains larger than $a_\mathrm{blow}$.

Collisions may play a significant role in the lifetimes of grains in the inner disk, as fragments from a catastrophically dispersed grain can be rapidly removed via radiation pressure.  For a low-eccentricity disk of single-size particles, the timescale between mutual collisions is $t_\mathrm{coll}\sim\left(\Omega\tau_\perp^\mathrm{geo}\right)^{-1}$, where $\Omega$ is the orbital frequency and $\tau_\perp^\mathrm{geo}$ is the geometric vertical optical depth.  We find $t_\mathrm{coll}\sim6\times10^4\left(r/60\,\mathrm{AU}\right)^{3/2}\left(\tau_\perp^\mathrm{geo}/10^{-3}\right)^{-1}$\,yr.  Calculation of this timescale from our models is complicated by the unknown absorption efficiency, as $\tau_\perp^\mathrm{geo}(\varpi_0)=\tau_0/Q_\mathrm{abs}$.  We also note that the timescale for \emph{destructive} collisions is likely different from the above $t_\mathrm{coll}$, and it must account for the unknown grain size and velocity distributions.  The presence of a gas disk can damp relative velocities, increasing the timescale for destructive collisions.  These damping forces can also cause grains to settle in the midplane, causing the disk to be very thin in vertical extent~\citep[e.g.][]{garaud_etal04}.  Observations of the disk scale height may constrain the presence of gas and its effect on collisional timescales.  Our models in~\S\ref{subsec:modeling} show that the mid-IR image is consistent with an edge-on, vertically unresolved disk.  More work is needed in modeling scattered-light images (at higher spatial resolution) to determine if the apparent disk thickness is the result of vertical extent or inclination effects.  Alternately, it may be possible to constrain the gas density through modeling of the dust density's radial structure, as has been done for $\beta$\,Pic~\citep{thebault&augereau05}.  Regardless, future dynamical models of the inner disk must make a detailed accounting grain-grain collisions.

Drag forces can decrease the periastra of grain orbits, filling the inner disk with material.
Assuming circular orbits, the Poynting-Robertson drag timescale at radius $r$ is $t_\mathrm{PR}\sim10^6\left(a/1\,\micron\right)\left(\rho/2\,\mathrm{g}\,\mathrm{cm}^{-3}\right)\left(r/60\,\mathrm{AU}\right)^2\left(Q_\mathrm{rad}/2\right)^{-1}$\,yr.  This is significantly longer than $t_\mathrm{coll}\lesssim 10^5$\,yr, suggesting P-R drag is dynamically unimportant for the radii of warm dust emission~\citep[though it may allow some mid-IR emitting dust in systems without planets; see][]{wyatt05}.
The detection of a potentially massive gas disk by~\citet{redfield07} suggests that gas drag may affect the grains in this system.  As noted above, the role of gas drag in the inner disk is difficult to estimate due to our ignorance of the gas disk's density distribution and physical state, and no physical features (such as sharp outer edges of dust rings or midplane settling) currently provide such indications~\citep[e.g.][]{takeuchi&artymowicz01, besla&wu07}.

An interesting physical result from our models are the warm temperatures of the effective grains at the inner rim of the dust annulus [$T(\varpi_0)\simeq130$--200\,K].  At these temperatures, the sublimation of water ice in small grains is efficient ($t_\mathrm{sub}\lesssim1$\,yr), suggesting no water is present in grains at this distance from the star.  However, this process is a strong function of temperature~\citep[cf. Eq. 16 of][]{backman&paresce93}, and therefore stellar distance (Eq.~\ref{eq:temp}).  At a distance of $r=2\varpi_0$, a 0.2\,\micron\ water ice grain has $t_\mathrm{sub}\sim1$\,Myr.  This raises the possibility that, rather than being the location of grain creation and outward diffusion, the inner edge of the warm dust disk is the destruction site for icy inspiraling grains.  For this to be the case, drag forces must overcome radiation pressure and destructive collisions for a significant population of icy grains.

The inference of the physical processes responsible for the disk structure (and that govern its evolution) are unclear.
Direct spectroscopic evidence of the stellar age is still needed.
Furthermore, the spatial distribution of the gas disk is an important direction for future observations.  This system is attractive for the direct detection of gas emission, similar to observations of $\beta$\,Pic~\citep{thi_etal01, olofsson_etal01, brandeker_etal04}.
The next modeling steps should combine the available data, including the resolved scattered light, thermal emission, and SED.
More detailed calculations can reveal the processes responsible for the disk structure, in a manner similar to that developed for AU\,Mic by~\citet{strubbe&chiang06}.

\subsection{Conclusions}\label{subsec:conclusions}

We have
(1) spatially resolved the thermal emission from the warm inner disk around HD\,32297 in the \bNp\ band,
(2) found that the stellar SED is inconsistent with the temperature and luminosity of A0; rather, we favor a cooler, less-luminous star,
(3) found that the observed \bNp\ emission with the stellar PSF subtracted suggests a symmetric, optically thin ring model,
(4) modeled the thermal emission (including \textit{IRAS} SED data) with an annulus consisting of a single population of efficiently absorbing, inefficiently radiating grains,
(5) determined that a separate population producing the 60\,\micron\ emission improves the fit to the SED and the consistency with the scattered light disk, and
(6) identified the possibility that ice sublimation may play a significant role in the destruction of grains in the warm inner disk.

\acknowledgements
We would like to thank Scott Fisher and Kevin Volk for their helpful observing support.
This publication makes use of data products from the Two Micron All Sky Survey, which is a joint project of the University of Massachusetts and the Infrared Processing and Analysis Center/California Institute of Technology, funded by NASA and the NSF.
Based on observations obtained at the Gemini Observatory, which is operated by AURA, Inc., under a cooperative agreement with the NSF on behalf of the Gemini partnership.
The authors acknowledge the significant cultural role that the summit of Mauna Kea has always had within the indigenous Hawaiian community.
This work was supported in part by the NSF Science and Technology Center for Adaptive Optics, managed by the University of California at Santa Cruz under cooperative agreement AST-9876783.

\bibliography{}

\end{document}